\documentclass[aps,pra,twocolumn,showpacs,superscriptaddress]{revtex4}
\usepackage{graphicx,dcolumn,bm,amsmath,amssymb,amsfonts}
\usepackage{graphicx}
\usepackage{amssymb,bm}
\begin{document}
\title{
Decrease of entanglement by local operations in the D\"ur--Cirac method
}
\author{Yukihiro Ota}
\email[Electronic address: ]{oota@hep.phys.waseda.ac.jp}
\affiliation{Department of Physics, Waseda University, Tokyo 169--8555,
Japan} 
\affiliation{Advanced Research Institute for Science and Technology, Waseda
University, Tokyo 169--8555, Japan}
\author{Motoyuki Yoshida}
\email[Electronic address: ]{motoyuki@hep.phys.waseda.ac.jp}
\affiliation{Department of Physics, Waseda University, Tokyo 169--8555,
Japan} 
\author{Ichiro Ohba}
\email[Electronic address: ]{ohba@waseda.jp}
\affiliation{Department of Physics, Waseda University, Tokyo 169--8555,
Japan}
\affiliation{Advanced Research Institute for Science and Technology, Waseda
University, Tokyo 169--8555, Japan}
\affiliation{Kagami Memorial Laboratory for Material Science and
Technology, Waseda University, Tokyo 169--0051, Japan}

\date{\today}
\begin{abstract}
One cannot always obtain information about entanglement by
 the D\"ur--Cirac (DC) method. 
The impracticality is attributed to the decrease of entanglement
 by local operations in the DC method. 
We show that, even in $2$--qubit systems, there exist states whose
 entangled property the DC method never evaluates. 
The class of such states in $2$--qubit systems is completely
 characterized by the value of the fully entangled fraction.  
Actually, a state whose fully entangled fraction is less than or
 equal to $\frac{1}{2}$ is always transformed into a separable state by
 local operations in the DC method, even if it has negative 
 partial transposition.  
\end{abstract}
\maketitle
\section{Introduction}
Quantum mechanics has a quite different mathematical and conceptual
structure from that of classical mechanics. 
Quantum entanglement vividly illustrates this point\,\cite{Peres1995}.   
Investigation into the character of entanglement is necessary for not
only the deep understanding of quantum theory but also its application. 
Indeed, entanglement is regarded as a key concept of quantum
information processing\,\cite{BEZ}.

The classification and quantification of bipartite entanglement (i.e.,
entanglement between two subsystems in a total quantum system) are well 
established\,\cite{Peres,HHH1996,Horodecki1997,HHH1998,BDSWW1996,VPRK,VP1998,ABHHHRWZ}.  
In particular, the positive partial transposition criterion (PPT)
\,\cite{Peres,HHH1996,Horodecki1997,HHH1998} is very useful,   
because one can readily obtain a sufficient condition for an entangled
state, a necessary condition for a separable state (i.e., a state with
no quantum correlation)\,\cite{Werner}, or a necessary condition for a
distillable state\,\cite{ABHHHRWZ}, by linear algebra. 

The situation becomes more complicated as the number of subsystems
in a total system increases. 
In 3--qubit systems, for example, there are two inequivalent classes
of entanglement.
By stochastic local operations and classical
communication\,\cite{BPRST2001}, a Greenberger--Horn--Zeilinger (GHZ)
state cannot be transformed into a $W$ state, and vice
versa\,\cite{DVC2000}. 
However, multiparticle entanglement can play an important role in
quantum protocol (e.g., quantum telecloning\,\cite{MJPV1999}) and
quantum computing. 
Moreover, its classification will be useful for deeply understanding
quantum phase transitions in condensed matter physics\,\cite{OAFF2002}. 
Thus, research into multiparticle entanglement is a crucial 
and popular issue in both quantum physics and quantum information theory.

Various attempts to classify and quantify multiparticle entanglement
have been made\,\cite{DCT,DC,DC62,DC2001,Miyake2003,CK_I,CK_II,IP2005}.
Among them, D\"ur and Cirac\,\cite{DC} proposed a systematic way of
classifying multiparticle entanglement in $N$--qubit systems. 
Hereafter, we call it the D\"ur--Cirac (DC) method.  
The main idea is that, using a sequence of local operations, one can
transform an arbitrary density matrix of an $N$--qubit system into
a state whose entangled property is easily examined. 
It should be noted that entanglement cannot increase through local
operations.  
Accordingly, if the density matrix transformed by local operations
is entangled, then the original density matrix represents an entangled
state. 

However, one cannot always obtain an entangled property by the DC
method.  
In our previous paper\,\cite{OMYO}, we suggested that there exists an
impracticality in the DC method through an example. 

In this letter, we reveal the possibility that one cannot obtain the
desired information on entanglement by the DC method, though it is a very
simple and effective method for examining multiqubit entanglement. 
We show that there is such a possibility even in $2$--qubit systems.  
The most important quantity in our discussion is a fully
entangled fraction\,\cite{BDSWW1996,ABHHHRWZ}. 
Our main result is that, in $2$--qubit systems, one can never determine
whether a quantum state is entangled or not through the DC method if the
fully entangled fraction is less than or equal to $\frac{1}{2}$. 
Then, we completely characterize the class of the state in $2$--qubit systems
whose entangled property is never obtained by the DC method.
The impracticality of the DC method is due to the decrease of
entanglement by local operations in the DC method. 
Additionally, we investigate what parts of the local operations reduce
the entanglement in $2$--qubit system. 

The letter is organized as follows.
We briefly review the DC method in section \ref{sec:review_DC}. 
Then, we illustrate the impracticality of the DC method through an example
in a $2$--qubit system, and show the relation to the local operations in 
section \ref{sec:localop}. 
After that, we show our main results in section \ref{sec:main}. 
Our results are shown only in $2$--qubit systems, but they clearly reveal the
limitation of the DC method. 
Section \ref{sec:summary} is devoted to a summary.

\section{Review of the DC method}
\label{sec:review_DC}
We briefly review the DC method\,\cite{DCT,DC,DC62,DC2001}. 
Its main idea is that, using a sequence of local operations, one can
transform an arbitrary density matrix of an $N$--qubit system into
a state whose property of entanglement is easily examined. 

First, we explain how to specify a bipartition of the system
concerned. 
We divide an $N$--qubit system into two subsystems, system A and system
B, as follows. 
Let us consider a set of binary numbers,
$\{k_{i}\}_{i=1}^{N}$ ($k_{i}=0,\,1$).  
When $k_{i}$ is equal to $0\,(1)$, the $i$th qubit is in system A (B). 
We always set $k_{1}=0$; the first qubit is always in system A. 
Representing the number by a binary,  $k\equiv \sum_{i=2}^{N}k_{i}2^{i-2}$, 
a partition is specified in the $N$--qubit system if an
integer $k(\in[1,\,2^{N-1}-1])$ is chosen; we call such a partition the
bipartition $k$. 

The authors in Ref.\,\cite{DCT,DC,DC62,DC2001} introduced a special
family of density matrices as follows: 
\begin{eqnarray}
\rho_{N}
&=&
\lambda_{0}^{+}
|\Psi_{0}^{+}\rangle\langle\Psi_{0}^{+}| 
+
\lambda_{0}^{-}
|\Psi_{0}^{-}\rangle\langle\Psi_{0}^{-}|
\nonumber \\
&&
\quad
+
\sum_{j=1}^{2^{N-1}-1}\lambda_{j}
\big(
|\Psi_{j}^{+}\rangle\langle\Psi_{j}^{+}|
+
|\Psi_{j}^{-}\rangle\langle\Psi_{j}^{-}|
\big),
\label{eq:ds} 
\end{eqnarray} 
where the coefficients $\lambda^{\pm}_{0}$ and $\lambda_{j}$ are real
and positive, and 
\(
\lambda^{+}_{0}+\lambda^{-}_{0}+\sum_{j=1}^{2^{N-1}-1}\lambda_{j}
=
1
\) 
because ${\rm tr}\rho_{N}=1$. 
These coefficients are related to the information on an arbitrary
density matrix of an $N$--qubit system, as shown below. 
The generalized GHZ state\,\cite{DCT,DC,DC62,DC2001} in an $N$--qubit
system 
\(
|\Psi^{\pm}_{j}\rangle 
\) is defined as follows:  
\begin{equation}
|\Psi^{\pm}_{j}\rangle 
=
\frac{1}{\sqrt{2}}
\left(
|0j\rangle \pm |1\bar{\jmath}\rangle
\right)\quad
(0\le j\le 2^{N-1}-1),
\label{eq:g_GHZ}
\end{equation}
where 
\(
j
\equiv 
\sum_{i=2}^{N}j_{i}2^{i-2}
\) 
for the binary number $j_{i}$ ($=0,\,1$), 
\(
|0j\rangle 
\equiv 
|0\rangle_{1}\otimes\bigotimes_{i=2}^{N}|j_{i}\rangle_{i}
\) and 
\(
|1\bar{\jmath}\rangle 
\equiv 
|1\rangle_{1}\otimes\bigotimes_{i=2}^{N}|1-j_{i}\rangle_{i}
\). 
The symbol $\bar{\jmath}$ means a bit--flip of $j$:
$\bar{\jmath}=2^{N-1}-1-j$. 
We write the computational basis for the $i$th qubit as 
$|0\rangle_{i}$ and $|1\rangle_{i}$ ($\,_{i}\langle 0|0\rangle_{i}=1$, 
$\,_{i}\langle 1|1\rangle_{i}=1$, and 
$\,_{i}\langle 0|1\rangle_{i}=0$). 
The subscription $i(=1,\,2,\ldots N)$ is the label of the qubit.
We can easily find the generalized GHZ states are the elements of an
orthonormal basis of the Hilbert space corresponding to the $N$--qubit
system. 
Note that the convention of generalized GHZ states\,(\ref{eq:g_GHZ}) is
slightly different from the corresponding one in
Ref.\,\cite{DCT,DC,DC62,DC2001}, but such a difference doesn't matter
in our discussion.

We summarize the several useful properties of $\rho_{N}$. 
The compact consequences for partial transposition with respect
to any bipartition are known\,\cite{DCT,DC,DC62,DC2001}.
First, $\rho_{N}$ has positive partial transposition (PPT) with respect
to a bipartition $k$ if and only if  
\(
\Delta \le 2\lambda_{k}
\), 
where 
\(
\Delta = |\lambda^{+}_{0}-\lambda^{-}_{0}|
\). 
On the other hand, $\rho_{N}$ has negative partial
transposition (NPT) with respect to a bipartition $k$ if and only if 
\(
\Delta > 2\lambda_{k}
\). 
Furthermore, the authors in Ref.\,\cite{DC,DC62} proved the theorems about
multiparticle entanglement.  
Among them, we explain an important one\,\cite{DC}.  
We concentrate on two qubits, for example the $i$th and $j$th
qubits, in an $N$--qubit system. 
Let us consider all possible bipartitions, $\mathcal{P}_{ij}$ under which
the $i$th and $j$th qubits belong to different parties. 
The theorem is that $\rho_{N}$ has NPT with respect to
$~^{\forall}k\in\mathcal{P}_{ij}$ if and only if the maximal entangled
states between the $i$th and $j$th qubits can be distilled. 

The most important result in Ref.\,\cite{DCT,DC} is that an
arbitrary density matrix, $\rho$  of an $N$--qubit system, can be
transformed into $\rho_{N}$ by local operations, and local operations
cannot increase entanglement.  
Accordingly, if $\rho_{N}$ is an entangled state with respect to a
bipartition, $\rho$ is also such a state. 
Moreover, according to the theorem explained at the end of the above
paragraph, if $\rho_{N}$ has NPT with respect to
$~^{\forall}k\in\mathcal{P}_{ij}$, the maximal entangled state between
the $i$th and $j$th qubits can be distilled from $\rho_{N}$.  
Then, one should be able to distill the maximal entangled state between
such qubits from $\rho$. 
This result implies that one can know the sufficient condition for
the distillability of $\rho$ for an arbitrary $N$.  
Note that, through the PPT criterion, one can only obtain the necessary
condition for the distillability in an $N$-qubit system when
$N>2$\,\cite{HHH1998}. 

Under the local operations, the coefficients $\lambda^{\pm}_{0}$ and
$\lambda_{j}$ of $\rho_{N}$ are given by the following relations: 
\begin{equation}
\lambda^{\pm}_{0} 
= 
\langle\Psi^{\pm}_{0}|\rho|\Psi^{\pm}_{0}\rangle, 
\quad
2\lambda_{j}
=
\langle\Psi^{+}_{j}|\rho|\Psi^{+}_{j}\rangle 
+
\langle\Psi^{-}_{j}|\rho|\Psi^{-}_{j}\rangle. 
\label{eq:lambda_def}
\end{equation}
Consequently, one can systematically treat the evaluation of
multiparticle entanglement as a task for bipartite entanglement,
because it is only necessary to calculate some specific matrix elements
of $\rho$. 
In addition, this point will be useful for investigating entanglement in
experiments\,\cite{DC2001}.
\section{Local operations in the DC method}
\label{sec:localop}
As shown in the previous section, one can readily evaluate the information
of multiparticle entanglement by the DC method. 
However, the desired information about entanglement isn't always
obtained.  
Let us illustrate such an impractical case by an example in a 2--qubit
system. 
One can easily find that, by the PPT criterion, the
following density matrix has NPT (i.e., entangled): 
\begin{eqnarray}
\rho_{f}
&=&
\frac{1}{2}|\Psi^{+}_{0}\rangle\langle\Psi^{+}_{0}|
+
\frac{1}{4}|\Psi^{+}_{1}\rangle\langle\Psi^{+}_{1}| 
\nonumber \\
&&
\,
+
\frac{1}{4}|\Psi^{-}_{1}\rangle\langle\Psi^{-}_{1}| 
+
\frac{1}{4}
\big(
|\Psi^{+}_{1}\rangle\langle\Psi^{-}_{1}|
+
|\Psi^{-}_{1}\rangle\langle\Psi^{+}_{1}|
\big).
\label{eq:ex1}
\end{eqnarray}
One needs only to calculate $\Delta$ and $2\lambda_{1}$ to apply the
DC method to a 2--qubit system. 
According to Eq.\,(\ref{eq:lambda_def}), one
can readily obtain the following results for $\rho_{f}$:
$\Delta=\frac{1}{2}$ and $2\lambda_{1}=\frac{1}{2}$. 
Then, it is not possible to determine whether $\rho_{f}$ is entangled or
not, because $\Delta=2\lambda_{1}$.

We will show that the above problem should be attributed to the decrease of
entanglement by local operations in the DC method. 
Let us explain D\"ur and Cirac's explicit expressions to clarify this point. 
The local operations in the DC method are sequence of the following
three steps.  
First, we perform the following probabilistic unitary operator on an
arbitrary density operator of an $N$--qubit system: 
\begin{equation}
\mathcal{L}_{1}\rho 
=
\frac{1}{2}\rho 
+
\frac{1}{2}W_{1}\rho W_{1}^{\dagger}, 
\label{eq:L1}
\end{equation}
where 
\(
W_{1} = \bigotimes_{i=1}^{N}\sigma_{x}^{(i)}
\) and 
\(
\sigma_{x}^{(i)} 
= 
|0\rangle_{i}\langle 1| + |1\rangle_{i}\langle 0|
\). 
Note that 
\begin{eqnarray}
 \rho 
&=&
\sum_{j,\,j^{\prime}=0}^{N}
\bigg(
\mu_{jj^{\prime}}^{++}
|\Psi^{+}_{j}\rangle\langle\Psi^{+}_{j^{\prime}}|
+
\mu_{jj^{\prime}}^{+-}
|\Psi^{+}_{j}\rangle\langle\Psi^{-}_{j^{\prime}}|
\nonumber \\
&&
\qquad\qquad
+
\mu_{jj^{\prime}}^{-+}
|\Psi^{-}_{j}\rangle\langle\Psi^{+}_{j^{\prime}}|
+
\mu_{jj^{\prime}}^{--}
|\Psi^{-}_{j}\rangle\langle\Psi^{-}_{j^{\prime}}|
\bigg), 
\end{eqnarray}
where $\mu_{jj^{\prime}}^{\sigma\sigma^{\prime}}$s are the matrix elements of $\rho$ for
the generalized GHZ states ($\sigma,\,\sigma^{\prime}=\pm$).
As a result of this operation, the terms corresponding to 
\(
|\Psi^{+}_{j}\rangle\langle\Psi^{-}_{j^{\prime}}|
\)
and 
\(
|\Psi^{-}_{j}\rangle\langle\Psi^{+}_{j^{\prime}}|
\)
are vanishing because 
\(
W_{1}|\Psi_{j}^{\pm}\rangle 
= 
\pm 
|\Psi_{j}^{\pm}\rangle
\).

The following probabilistic unitary operators are necessary for the
second step: 
\begin{equation}
\mathcal{L}_{l}\rho 
=
\frac{1}{2}\rho
+ 
\frac{1}{2}W_{l}\rho W_{l}^{\dagger}
\qquad
(l=2,\,3,\,\ldots,\,N),
\label{eq:Lp}
\end{equation}
where
\(
W_{l} = \sigma^{(1)}_{z}\otimes\sigma^{(l)}_{z}
\) and 
\(
\sigma^{(i)}_{z} 
= 
|0\rangle_{i}\langle 0|
-
|1\rangle_{i}\langle 1|
\).
Equation\,(\ref{eq:Lp}) is a local operation with respect to the first
and $l$th qubit. 
Note that we abbreviate the identity operators for the other qubits in
$W_{l}$. 
In the second step, we perform 
\(
\prod_{l=2}^{N}\mathcal{L}_{l}
\) 
on the result of the first step. 
By this operation, the terms corresponding to 
\(
|\Psi^{\pm}_{j}\rangle\langle\Psi^{\pm}_{j^{\prime}}|
\)
($j\neq j^{\prime}$) are vanishing because 
\(
W_{l}|\Psi^{\pm}_{j}\rangle = (-1)^{j_{l}}|\Psi^{\pm}_{j}\rangle
\).
In this stage, the resultant state is a diagonal form with respect to
the generalized GHZ states. 
	
Finally, we perform the local random phase--shift,
$\mathcal{L}_{r}$ on the result of the second step: 
\begin{equation}
\mathcal{L}_{r}\rho
=
\prod_{i=1}^{N}
\bigg(\int^{2\pi}_{0}\frac{d\phi_{i}}{2\pi}\bigg)
2\pi \,\delta(\Phi-2\pi)\,
R_{\phi}\,\rho\,R_{\phi}^{\dagger}, 
\label{eq:Llr} 
\end{equation} 
where 
\(
R_{\phi}
=
\bigotimes_{i=1}^{N}R^{(i)}(\phi_{i})
\), 
\(
R^{(i)}(\phi_{i})|0\rangle_{i} = e^{i\phi_{i}}|0\rangle_{i}
\), 
\(
R^{(i)}(\phi_{i})|1\rangle_{i} = |1\rangle_{i}
\), 
and 
\(
\Phi = \sum_{i=1}^{N}\phi_{i}
\). 
Note that 
\(
\mathcal{L}_{r}\,
|\Psi^{\pm}_{0}\rangle\langle\Psi^{\pm}_{0}|
=
|\Psi^{\pm}_{0}\rangle\langle\Psi^{\pm}_{0}|
\) 
and 
\(
\mathcal{L}_{r}\,
|\Psi^{\pm}_{j}\rangle\langle\Psi^{\pm}_{j}|
=
\frac{1}{2}
(
|0j\rangle\langle 0j| 
+ 
|1\bar{\jmath}\rangle\langle 1\bar{\jmath}|
)
\) 
($j\neq 0$).  
After the final step, we can find that the resultant state is equivalent
to Eq.\,(\ref{eq:ds}).

Now, let us go back to Eq.\,(\ref{eq:ex1}). 
We only need to perform $\mathcal{L}_{1}$ on $\rho_{f}$ to transform it into
the form of Eq.\,(\ref{eq:ds}): 
\(
\mathcal{L}_{1}\rho_{f} 
=
\frac{1}{2}|\Psi^{+}_{0}\rangle\langle\Psi^{+}_{0}|
+
\frac{1}{4}|\Psi^{+}_{1}\rangle\langle\Psi^{+}_{1}|
+
\frac{1}{4}|\Psi^{-}_{1}\rangle\langle\Psi^{-}_{1}|
\). 
Obviously, the resultant state is separable. 
It implies that the entanglement decreases by the local operation
$\mathcal{L}_{1}$. 
In the subsequent section, we will characterize the class of the quantum
states in a $2$--qubit system whose entangled property is not
obtained by the DC method due to its decrease by the local operations.
\section{Limitation of the DC method in $2$--qubit systems}
\label{sec:main}
We attempt to reveal the class of the quantum states whose entangled
property is not obtained by the DC method.  
In this section, we focus on the case $N=2$ because its entanglement
structure is well known. 

Let us first introduce an important quantity for our
consideration: 
\begin{equation}
\mathcal{F}(\rho) 
= 
\max_{U,\,V}\,
\langle\Psi^{+}_{0}| 
(U\otimes V)\,\rho\,(U\otimes V)^{\dagger}
|\Psi^{+}_{0}\rangle,
\label{eq:full_ef}
\end{equation}
where $U$ and $V$ are unitary operators on the Hilbert
spaces for the first and second qubits, respectively. 
Equation (\ref{eq:full_ef}) is called a fully entangled
fraction\,\cite{BDSWW1996,ABHHHRWZ}. 

We show that the value of a fully entangled fraction plays an important
role in determining whether the DC method works or not. 
According to the DC method, the sufficient condition for an entangled state
in a $2$--qubit system is $\Delta > 2\lambda_{1}$.  
Using ${\rm tr}\rho=1$ and Eq.\,(\ref{eq:lambda_def}), we readily
obtain the following relation:  
\begin{eqnarray}
\Delta > 2\lambda_{1}
\iff 
\langle\Psi^{+}_{0}|\rho|\Psi^{+}_{0}\rangle
>\frac{1}{2}
\quad
{\rm or}
\quad
\langle\Psi^{-}_{0}|\rho|\Psi^{-}_{0}\rangle
>\frac{1}{2}.
\label{eq:DCNPT_fraction}
\end{eqnarray}
The right--hand side of Eq.\,(\ref{eq:DCNPT_fraction}) implies
$\mathcal{F}(\rho)>\frac{1}{2}$.
Note that 
\(
|\Psi^{-}_{0}\rangle = (I^{(1)}\otimes\sigma^{(2)}_{z})|\Psi^{+}_{0}\rangle
\), 
where $I^{(1)} = |0\rangle_{1}\langle 0|+|1\rangle_{1}\langle 1|$. 
Summarizing the above argument, we obtain the following statements: 
\begin{equation}
\Delta >2\lambda_{1}\,
\Longrightarrow\,
\mathcal{F}(\rho) > \frac{1}{2},
\label{eq:state1}
\end{equation}
or 
\begin{equation}
\mathcal{F}(\rho) \le \frac{1}{2}\,
\Longrightarrow \,
\Delta \le 2\lambda_{1}.
\label{eq:state2} 
\end{equation}
Accordingly, we obtain the following conclusion. 
Let us consider the density matrix in a $2$--qubit system which has NPT;
it is an entangled state. 
However, if its fully entangled fraction is less than or equal to
$\frac{1}{2}$, then it is not possible to determine whether such a state
is entangled or not by the DC method. 
Actually, Eq.\,(\ref{eq:ex1}) is just such an example. 

Next, we investigate the density matrix $\rho$ whose fully entangled
fraction is greater than $\frac{1}{2}$. 
In general, the condition $\mathcal{F}(\rho)>\frac{1}{2}$ does not imply 
\(
\langle\Psi^{\pm}_{0}|\rho|\Psi^{\pm}_{0}\rangle
>\frac{1}{2}
\). 
However, the following statement is always true: 
\begin{equation}
~^{\exists}\tilde{U}\otimes \tilde{V} \quad
{\rm s.t.}\quad
|\Psi^{+}_{0}\rangle 
=
\tilde{U}\otimes \tilde{V} |\tilde{\psi}\rangle,  
\end{equation}
where $\tilde{U}$ and $\tilde{V}$ are unitary operators on the Hilbert
spaces for the first and second qubits, respectively, and
$|\tilde{\psi}\rangle$ is the maximally entangled state that satisfies
\(
\langle\tilde{\psi}|\rho|\tilde{\psi}\rangle 
=
\mathcal{F}(\rho)
\). 
Consequently, using the above local unitary operator, we obtain 
\begin{eqnarray}
&&
\langle\Psi^{+}_{0}|\tilde{\rho}|\Psi^{+}_{0}\rangle > \frac{1}{2}, 
\label{eq:tilderho_frac} \\
&&
\tilde{\rho} 
=
(\tilde{U}\otimes\tilde{V})\,
\rho\,
(\tilde{U}\otimes\tilde{V})^{\dagger}.
\label{eq:def_tilderho}
\end{eqnarray}
According to Eqs.\,(\ref{eq:state1}) and (\ref{eq:tilderho_frac}), we
obtain the following statement: 
\begin{equation}
\mathcal{F}(\rho)>\frac{1}{2}\,
\Longrightarrow\,
~^{\exists}\tilde{U}\otimes\tilde{V}\quad
{\rm s.t.}\quad
\tilde{\Delta}
\equiv|\tilde{\lambda}^{+}_{0}-\tilde{\lambda}^{-}_{0}| 
>2\tilde{\lambda}_{1}, 
\label{eq:state3}
\end{equation}
where \(
\tilde{\lambda}^{\pm}_{0} 
= 
\langle\Psi^{\pm}_{0}|\tilde{\rho}|\Psi^{\pm}_{0}\rangle
\)
and 
\(
2\tilde{\lambda}_{1}
=
\langle\Psi^{+}_{1}|\tilde{\rho}|\Psi^{+}_{1}\rangle 
+
\langle\Psi^{-}_{1}|\tilde{\rho}|\Psi^{-}_{1}\rangle
\). 
The local unitary transformed state $\tilde{\rho}$ is entangled if 
\(
\tilde{\Delta}>2\tilde{\lambda}_{1}
\); one can obtain the entangled property of $\tilde{\rho}$ by the
DC method.   
On the other hand, the original density matrix $\rho$ is related to
$\tilde{\rho}$ through the local unitary operator
$\tilde{U}\otimes\tilde{V}$ from Eq.\,(\ref{eq:def_tilderho});
$\tilde{\rho}$ is equivalent to $\rho$ with respect to entanglement. 
Therefore, one can obtain the entangled property of a density matrix
whose fully entangled fraction is greater than $\frac{1}{2}$  by the DC
method with a suitable local unitary operator.    
Let us show an example for such a case. 
We consider a Bell--diagonal state. 
Such a state is defined by as follows: 
\begin{equation}
\rho_{{\rm BD}}
=
\sum_{j=0}^{1}
\left(
\mu^{+}_{j}|\Psi^{+}_{j}\rangle\langle\Psi^{+}_{j}|
+
\mu^{-}_{j}|\Psi^{-}_{j}\rangle\langle\Psi^{-}_{j}|
\right),
\label{eq:bd_state} 
\end{equation}
where $\mu^{\pm}_{j} \ge 0$ and
$\sum_{j=0}^{1}(\mu^{+}_{j}+\mu^{-}_{j})=1$. 
Note that our example in Ref.\,\cite{OMYO} was a special case of
Eq.\,(\ref{eq:bd_state}). 
We can show that  
$\rho_{{\rm BD}}$ has NPT if and only if 
\begin{equation}
|\mu_{0}^{+}-\mu_{0}^{-}| > \mu_{1}^{+}+\mu_{1}^{-}
\quad
{\rm or} 
\quad
|\mu_{1}^{+}-\mu_{1}^{-}| > \mu_{0}^{+}+\mu_{0}^{-}.
\label{eq:NPT_BD_cond}
\end{equation}
According to ${\rm tr}\,\rho_{{\rm BD}}=1$ and
Eq.\,(\ref{eq:NPT_BD_cond}), if one of $\mu^{\sigma}_{j}$s
($\sigma=\pm$) is at least greater than $\frac{1}{2}$, then 
$\rho_{{\rm BD}}$ has NPT, and vice versa. 
In addition, we easily obtain the following relation: 
\begin{equation}
\mathcal{F}(\rho_{{\rm BD}}) 
=
\max_{\sigma=\pm,\,j=0,\,1}\, \mu^{\sigma}_{j}.
\label{eq:fraction_BD} 
\end{equation} 
Then, if $\rho_{{\rm BD}}$ is entangled, $\mathcal{F}(\rho_{{\rm BD}})$
is greater than $\frac{1}{2}$.  
In this case, we can obtain the information of the entanglement for the
Bell--diagonal state in a $2$--qubit system by the DC method with a suitable
local operator. 
Note that one only needs to use $\mathcal{L}_{r}$ to transform
$\rho_{{\rm BD}}$ into $\rho_{N}$. 

Finally, we consider whether, through the DC method with appropriate local
unitary operators,  we can obtain the entangled property of the quantum
state whose fully entangled fraction is less than or equal to $\frac{1}{2}$.  
It should be noted that the converse statement of
Eq.\,(\ref{eq:state3}) can be easily shown. 
Therefore, we conclude that one never obtains the entangled property for
a density matrix whose fully entangled fraction is less than or equal to
$\frac{1}{2}$ by the DC method, even if one uses local unitary
operators.  

In summary, we have completely classified the states in
$2$--qubit systems whose entangled property is not obtained by
the DC method, or by the DC method with local unitary operators. 
The limitation of the method is determined by the value of the fully
entangled fraction. 
If it is greater than $\frac{1}{2}$, we can always obtain the
desired information on entanglement by the DC method with suitable local
unitary operators. 
Otherwise, we never obtain it. 
The impracticality of the DC method is attributed to the decrease of
entanglement by the local operations.   
Note that the Bell--diagonal state is entangled if 
$\mathcal{F}(\rho_{{\rm BD}})>\frac{1}{2}$. 
Moreover, we can easily find 
\(
\mathcal{L}_{1}\,\rho_{{\rm BD}}
=
\mathcal{L}_{2}\,\rho_{{\rm BD}}
=
\rho_{{\rm BD}}
\) 
and 
$\mathcal{L}_{r}\rho_{{\rm BD}}=\rho_{N}$.
Accordingly, in $2$--qubit systems, the crucial decrease of entanglement
occurs in $\mathcal{L}_{1}$ and $\mathcal{L}_{2}$.  

\section{Summary}
\label{sec:summary}
We have shown that one cannot always obtain an entangled property by
the DC method, even in $2$--qubit systems. 
The most important quantity in our discussion is a fully entangled
fraction. 
One can never determine whether a quantum state is entangled or not
through the DC method, if the fully entangled fraction is less than or equal
to $\frac{1}{2}$. 
On the other hand, one can make such a determination by the DC method
with suitable local unitary operators, if the fully entangled fraction
is greater than  $\frac{1}{2}$. 

The impracticality of the DC method is attributed to the decrease of entanglement
by the local operations. 
Actually, from Eqs.\,(\ref{eq:ex1}) and (\ref{eq:L1}), we have easily
shown that 
\(
\mathcal{L}_{1}\rho_{f} 
\) 
is separable, even if $\rho_{f}$ is entangled.
The Bell--diagonal state (\ref{eq:bd_state}) is invariant under
$\mathcal{L}_{1}$ and $\mathcal{L}_{2}$; we only need to
use $\mathcal{L}_{r}$ for transforming it into the form of
Eq.\,(\ref{eq:ds}). 
In addition, the Bell--diagonal state which is entangled has a fully
entangled fraction greater than $\frac{1}{2}$. 
Therefore, the crucial decrease of entanglement for examining it by the DC
method occurs in $\mathcal{L}_{1}$ and $\mathcal{L}_{2}$ in $2$--qubit
systems.  

Finally, we would like to comment on the case of multiqubit systems. 
The DC method has been proposed as a systematic estimation of multiparticle
entanglement. 
Therefore, it is necessary to study the limitation of the method in $N$--qubit
systems when $N>2$. 
However, the situation will be more complicated in this case. 
Nevertheless, the results in this letter can hint at a solution.
Namely, we will consider the following question: 
(i) Is it possible to obtain the entangled property of Bell--diagonal
states in $N$--qubit systems, 
\begin{equation}
\rho_{{\rm BD}} 
= 
\sum_{j=0}^{2^{N-1}-1} 
\left(
\mu^{+}_{j}|\Psi^{+}_{j}\rangle\langle\Psi^{+}_{j}|
+
\mu^{-}_{j}|\Psi^{-}_{j}\rangle\langle\Psi^{-}_{j}|
\right), 
\label{eq:bd_state_N} 
\end{equation}
by the DC method with suitable local unitary operators? 
(ii) How are fragile quantum states with respect to entanglement, for
example $\rho_{f}$, under the local operations characterized in
$N$--qubit systems? 
We think the above questions are related to the decrease of quantum
entanglement under local operations and decoherence. 
In addition, our examination of the above questions will lead to the
understanding of the structure of quantum states in $N$--qubit systems. 
\begin{acknowledgements}
The authors acknowledge H. Nakazato for valuable discussions. 
This research is partially supported by a Grant--in--Aid for Priority Area B
 (No.\,763), MEXT, by the 21st Century COE Program (Physics of
 Self-Organization Systems) at Waseda University from MEXT, and by a
 Waseda University Grant for Special Research Projects
 (Nos.\,2004B--872 and 2007A--044).  
\end{acknowledgements}


\begin{thebibliography}{99}
\bibitem{Peres1995}
A. Peres, 
{\it Quantum Theory: Concepts and Methods} 
(Kluwer Academic Publishers, Dordrecht, 1995).
\bibitem{BEZ}
{\it The Physics of Quantum Information: Quantum Cryptography, Quantum
	Teleportation, Quantum Computation},  
edited by D. Bouwmeester, A. Ekert, and A. Zeilinger  
(Springer, Berlin, 2000). 
\bibitem{Peres}
A. Peres, 
Phys. Rev. Lett. {\bf 77}, 1413 (1996).
\bibitem{HHH1996}
M. Horodecki, P. Horodecki, and R. Horodecki, 
Phys. Lett. A {\bf 223}, 1 (1996). 
\bibitem{Horodecki1997} 
P. Horodecki, Phys. Lett. A {\bf 232}, 333 (1997).
\bibitem{HHH1998}
M. Horodecki, P. Horodecki, and R. Horodecki, 
Phys. Rev. Lett. {\bf 80}, 5239 (1998).
\bibitem{BDSWW1996}
C. H. Bennett, D. P. DiVincenzo, J. A. Smolin, and W. K. Wootters, 
Phys. Rev. A {\bf 54}, 3824 (1996).
\bibitem{VPRK} 
V. Vedral, M. B. Plenio, M. A. Rippin, and P. L. Knight, 
Phys. Rev. Lett. {\bf 78}, 2275 (1997). 
\bibitem{VP1998}
V. Vedral and M.B. Plenio, 
Phys. Rev. A {\bf 57}, 1619 (1998).
\bibitem{ABHHHRWZ}
G. Alber, T. Beth, M. Horodecki, P. Horodecki, R. Horodecki,
	M. R\"{o}tteler, H. Weinfurter, R. Werner, and A. Zeilinger, 
{\it Quantum Information: An Introduction to Basic Theoretical Concepts and
	Experiments} 
(Springer, Berlin, 2001).
\bibitem{Werner}
R. F. Werner, 
Phys. Rev. A {\bf 40}, 4277 (1989).
\bibitem{BPRST2001}
C. H. Bennett, S. Popescu, D. Rohrlich, J. A. Smolin, and A. V. Thapliyal, 
Phys. Rev. A {\bf 63}, 012307 (2001). 
\bibitem{DVC2000}
W. D\"ur, G. Vidal, and J. I. Cirac, 
Phys. Rev. A {\bf 62}, 062314 (2000). 
\bibitem{MJPV1999}
M. Murao, D. Jonathan, M. B. Plenio, and V. Vedral, 
Phys. Rev. A {\bf 59}, 156 (1999). 
\bibitem{OAFF2002}
A. Osterloh, L. Amico, G. Falci, and R. Fazio, 
Nature {\bf 416}, 608 (2002). 
\bibitem{DCT}
W. D\"ur, J. I. Cirac, and R. Tarrach, 
Phys. Rev. Lett. {\bf 83}, 3562 (1999).
\bibitem{DC}
W. D\"ur and J. I. Cirac, 
Phys. Rev. A {\bf 61}, 042314 (2000).
\bibitem{DC62} 
W. D\"ur and J. I. Cirac, 
Phys. Rev. A {\bf 62}, 022302 (2000). 
\bibitem{DC2001}
W. D\"ur and J. I. Cirac, 
J. Phys. A {\bf 34}, 6837 (2001).
\bibitem{Miyake2003}
A. Miyake, 
Phys. Rev. A {\bf 67}, 012108 (2003). 
\bibitem{CK_I}
D. Chru\'{s}ci\'{n}ski and A. Kossakowski, 
Phys. Rev. A {\bf 73}, 062314 (2006).
\bibitem{CK_II}
D. Chru\'{s}ci\'{n}ski and A. Kossakowski, 
Phys. Rev. A {\bf 73}, 062315 (2006).
\bibitem{IP2005}
S. Ishizaka and M. B. Plenio, 
Phys. Rev. A {\bf 71}, 052303 (2005).
\bibitem{OMYO}
Y. Ota, S. Mikami, M. Yoshida, and I. Ohba, 
quant--ph/0612158. 
\end{thebibliography}
\end{document}